\documentclass[manuscript]{aastex63}

\received{April 14, 2021}
\revised{July 16, 2021}
\accepted{August 5, 2021}
\submitjournal{and Accepted in the Planetary Science Journal}

\shorttitle{UD in the Spotlight}
\shortauthors{Kareta et al.}

\graphicspath{{./}{figures/}}

\begin{document}

\title{Investigating the Relationship between (3200) Phaethon and (155140) 2005 UD through Telescopic and Laboratory Studies}

\correspondingauthor{Theodore Kareta}
\email{tkareta@lpl.arizona.edu}

\author[0000-0003-1008-7499]{Theodore Kareta}
\affil{Lunar and Planetary Laboratory \\
University of Arizona \\
Tucson, AZ, USA}

\author{Vishnu Reddy}
\affil{Lunar and Planetary Laboratory \\
University of Arizona \\
Tucson, AZ, USA}

\author{Neil Pearson}
\affil{Planetary Science Institute \\
Tucson, AZ, USA}

\author{Juan A. Sanchez}
\affil{Planetary Science Institute \\
Tucson, AZ, USA}

\author{Walter M. Harris}
\affil{Lunar and Planetary Laboratory \\
University of Arizona \\
Tucson, AZ, USA}

\begin{abstract}

The relationship between the Near-Earth Objects (3200) Phaethon and (155140) 2005 UD is unclear. While both are parents to Meteor Showers, (the Geminids and Daytime Sextantids, respectively), have similar visible-wavelength reflectance spectra and orbits, dynamical investigations have failed to find any likely method to link the two objects in the recent past. Here we present the first near-infrared reflectance spectrum of 2005 UD, which shows it to be consistently linear and red-sloped unlike Phaethon's very blue and concave spectrum. Searching for a process that could alter some common starting material to both of these end states, we hypothesized that the two objects had been heated to different extents, motivated by their near-Sun orbits, the composition of Geminid meteoroids, and previous models of Phaethon's surface. We thus set about building a new laboratory apparatus to acquire reflectance spectra of meteoritic samples after heating to higher temperatures than available in the literature to test this hypothesis and were loaned a sample of the CI Chondrite Orgueil from the Vatican Meteorite Collection for testing. We find that while Phaethon's spectrum shares many similarities with different CI Chondrites, 2005 UD's does not. We thus conclude that the most likely relationship between the two objects is that their similar properties are only by coincidence as opposed to a parent-fragment scenario, though the ultimate test will be when JAXA's \textit{DESTINY+} mission visits one or both of the objects later this decade. We also discuss possible paths forward to understanding Phaethon's properties from dynamical and compositional grounds.

\end{abstract}


\section{Introduction} \label{sec:intro}
\subsection{Overview}
The Geminid Meteor Shower is one of the best-studied and easily observed meteor showers, but until 1983 no parent comet had been identified. Furthermore, the orbit of the Geminids is not that of a Long-Period or Jupiter-Family comet, with a period of only 1.4 years and a perihelion distance inside the orbit of Mercury at $q=0.140$ AU. Near-Earth object (NEO) 1983 TB, later named (3200) Phaethon, was discovered in a Geminid-like orbit and was quickly deemed a plausible parent for the meteor shower \citep{1983IAUC.3881....1W}. While Phaethon was originally presumed to be a dead comet (see, e.g., \citealt{1999AdSpR..24.1167B}), with the Geminids thought to have been produced through conventional cometary activity in the past, the story has become much less clear over the past two decades. While studies of the Geminids themselves have mostly concluded that they are fundamentally cometary in nature, studies of Phaethon have been converging on the idea that it is an (active) asteroid from the Main Belt as opposed to stochastically captured from the outer Solar System (a review of both of these topics is presented in the next subsection). Phaethon is thus a key object to understand on the now-recognized continuum between comets and asteroids. The NEO (155140) 2005 UD \citep{2005A&A...438L..17K, 2006AJ....132.1624J} is compelling for the same reasons, as it is the inactive parent of the Daytime Sextantids -- as well as proposed fragment of Phaethon itself. Investigations into the modern properties and relationship between members of the 'Phaethon Geminid Complex' (PGC) are thus investigations into how meteor showers form as well as the properties of low-albedo NEOs.

\subsection{(3200) Phaethon}
Phaethon's properties are much better studied than 2005 UD's due to its earlier discovery date and larger size and thus sets the framework for which these related bodies are compared and understood. The surface of Phaethon is strongly blue at visible and near-infrared wavelengths ($\sim0.35-2.5$ $\mu{m}$, see, e.g. \citealt{,1984PhDT.........3T, 2007A&A...461..751L}), unlike the typical strongly red reflectance spectrum of traditional cometary nuclei and D-type asteroids. Phaethon's orbit, while highly eccentric to the point where its perihelion is lower than Mercury's at $q=0.14 AU$, is generally accepted to be hard to produce from a typical Jupiter Family Comet orbit \citep{2002Icar..156..399B} and probably more likely to come from the Main Belt of asteroids \citep{2010A&A...513A..26D}, possibly from (2) Pallas's collisonal family. However, there are serious issues with producing Phaethon's reflectance spectrum and albedo from an originally Pallas-like composition (see the discussion of \citealt{2018AJ....156..287K}). Furthermore, the radar-derived size of Phaethon of more than $6 km$ across at the equator \citep{2019P&SS..167....1T} is inconsistent with much previous thermal modeling of the object, supporting a lower albedo nearer to $\sim8\%$ like that of \citealt{2018AJ....156..287K} as opposed to the higher $\sim12-16\%$ values of studies like that of \citealt{2016A&A...592A..34H} or \citealt{2019AJ....158...97M} for reasons not yet well understood, though many of these authors have speculated that Phaethon may have odd thermal properties in light of the discrepancy. While Phaethon's visible-wavelength albedo might be typical for B-type asteroids \citep{2013A&A...554A..71A}, its radar albedo might be more consistent with comets than B-types \citep{2019P&SS..167....1T}. Previous models of Phaethon's surface largely involved hydrated materials \citep{2007A&A...461..751L}, but the surface of Phaethon appears to be totally dehydrated \citep{2020NatCo..11.2050T}. All of these non-converging lines of inquiry about Phaethon's surface are further complicated by the revelation that Phaethon is not totally inactive, but actually develops a small dust tail near perihelion \citep{2010AJ....140.1519J,2013ApJ...771L..36J}. This dust tail is most likely caused by intense solar radiation pressure sweeping small ($\sim1\mu{m}$) dust grains off the surface, not sublimation of volatiles or another process, though Phaethon's short rotation period ($\sim3.604$ hours, \citealt{2016A&A...592A..34H}) may assist the process \citep{2020ApJ...892L..22N}. This is assumed to be a process that could happen on any sufficiently sungrazing object, though it has yet to be observed conclusively on other objects (see a discussion in \citealt{2016ApJ...823L...6K}).

\subsection{The Geminids}
As opposed to the strange properties of their mostly-inactive parent body, the study of the orbital distribution and physical properties of the Geminids themselves has been a cleaner story. The age of the Geminids appears to be among the youngest of the large meteor showers, with a maximum age of a few thousand years \citep{1999SoSyR..33..224R, 2016MNRAS.456...78R}. The spatial distribution of meteoroids, especially its asymmetry about the peak, probably requires a large change in parent body orbit during stream ejection, such as by non-gravitational forces \citep{1985AVest..19..152L, 2016MNRAS.456...78R}. The meteors themselves appear to be denser than typical cometary meteors \citep{1988Icar...76..279H}. They have a wide variety of sodium abundances \citep{2005A&A...438L..17K, 2010pim7.conf...42B, ABE2020105040}, probably due to intense solar heating \citep{2009EM&P..105..321K}, but have an overall composition consistent with cometary material \citep{2010pim7.conf...42B, ABE2020105040}. The higher-than-cometary densities could also be explained as a heating-related effect, with the weaker grains being preferentially removed from the stream with time by thermal stresses \citep{2012A&A...539A..25C}. \citealt{ABE2020105040} found that smaller meteoroids were more depleted in sodium, consistent with this idea.

\subsection{The Phaethon-Geminid-Complex and Paper Overview}
A comprehensive explanation of the 'Phaethon-Geminid-Complex' (PGC) needs to explain the modern asteroid-like properties of Phaethon (surface reflectivity, orbit) as well as the comet-like properties of the Geminid meteors themselves (composition, orbital distribution) as described in the previous section. One avenue for further study is characterizing other small bodies thought to be dynamically associated with Phaethon and the Geminids. (155140) 2005 UD is a $\sim 1.6$ km NEO with a blue reflectance spectrum \citep{2006AJ....132.1624J, 2020PSJ.....1...15D} thought to be dynamically linked to Phaethon and the Daytime Sextantids, as well as the Geminids less directly \citep{2005A&A...438L..17K}. This is a second NEO with similar (visible) reflectance properties, a very low perihelion ($q=0.16 AU$), and associated with its own meteor shower, and thus is expected to be a highly valuable target for comparison with Phaethon. 

The specifics of the relationship between the two bodies are debated. For example, \citealt{2019MNRAS.485.3378R} makes dynamical arguments that the two objects could not have been in close contact in the recent past given our understanding of their orbits despite their similar-and-rare surface properties. It is unclear whether or not 2005 UD (the so-called 'mini-Phaethon') is a fragment of its larger cousin and thus genetically related, introduced to near-Earth space through some common process, or that the two objects just coincidentally have similar orbits, surfaces, and associated meteor showers. It is imperative to understand the relationship between these objects not just to better understand the origin of their modern properties and meteor showers, but also to make preparations and predictions for the arrival of \textit{JAXA}'s $DESTINY^+$ spacecraft at Phaethon in the mid-2020s \citep{2018LPI....49.2570A}, which may be able to fly-by 2005 UD at a later date. There is another object proposed to be in the 'Phaethon-Geminid-Complex', 225416 (1999 YC) \citep{2008M&PSA..43.5055O}, but its orbit and physical properties are even more different from Phaethon than UD's \citep{2008AJ....136..881K}, and has received comparatively little characterization as a result.

In this paper, we present the first near-infrared reflectance spectra of (155140) 2005 UD in Section 2 to better understand its modern properties and compare it to its better studied cousin (3200) Phaethon. We also present new laboratory reflectance measurements of the CI chondrite Orgueil heated to Phaethon and 2005 UD relevant temperatures in Section 3 to better constrain the thermal histories of and relationship between these bodies. In Section 4, we discuss how these interrelated lines of evidence and inquiry change our perception of their respective properties and histories.

\section{Near-Infrared Observations of 2005 UD} \label{sec:obs}
\begin{deluxetable*}{cccccCrlcc}[b!]
\tablecaption{\label{tab:observations}}
\tablecolumns{9}
\tablenum{1}
\tablewidth{0pt}
\tablehead{
\colhead{UT Date\tablenotemark{a}} &
\colhead{UTC Time\tablenotemark{a}} &
\colhead{$R_H$} &
\colhead{$\Delta$} &
\colhead{$\alpha$} & 
\colhead{$m_V$} & 
\colhead{Airmass} & 
\colhead{Humidity} & 
\colhead{Total Time}\\      
\colhead{(YYYY-mm-dd)} & 
\colhead{(d)} &
\colhead{(au)} &
\colhead{(au)} &
\colhead{$(^{\circ})$} & 
\colhead{} & 
\colhead{} & 
\colhead{} & 
\colhead{seconds}
}
\startdata
2018-09-19 & 14:39 & 0.98 & 0.27 & 86.6 & 17.5 & $\sim$1.26 & $\sim$26\% & 800.0  \\
2018-10-02 & 12:27 & 1.18 & 0.23 & 34.2 & 16.0 & 1.03-1.06 & 9-14\%  & 2800.0  \\
2018-10-03 & 09:12 & 1.20 & 0.23 & 30.5 & 16.0  &1.20-1.60 & 5-6\% & 5600.0  \\
\enddata
\tablenotetext{a}{At observing slot start, see text for specific details.}
\end{deluxetable*}

We obtained the first near-infrared reflectance spectra of 2005 UD at the NASA IRTF using the SpeX instrument \citep{rayner_spex:_2003} on three dates in September and October 2018, the details of which are presented in Table 1. Guiding was accomplished with the MORIS camera \citep{2011PASP..123..461G} due to the high apparent motion and faintness of the target. The observations followed the procedure of \citealt{2018AJ....156..287K}, whereby observations of the target were 'bookended' by observations of a local G-type star for immediate telluric correction and later slope-corrected further using a well-characterized solar analog G2V star (SAO 93936) observed near-zenith. This method has been used to effectively and reproducibly characterize faint Solar System small bodies on many occasions (see, e.g., \citealt{2019AJ....158..204S}.) The reduction was performed partially within the 'spextool' environment \citep{2004PASP..116..362C} and partially within custom-written scripts in Python (see 'Software' after the Acknowledgements section).

\begin{figure}[ht!]
\plotone{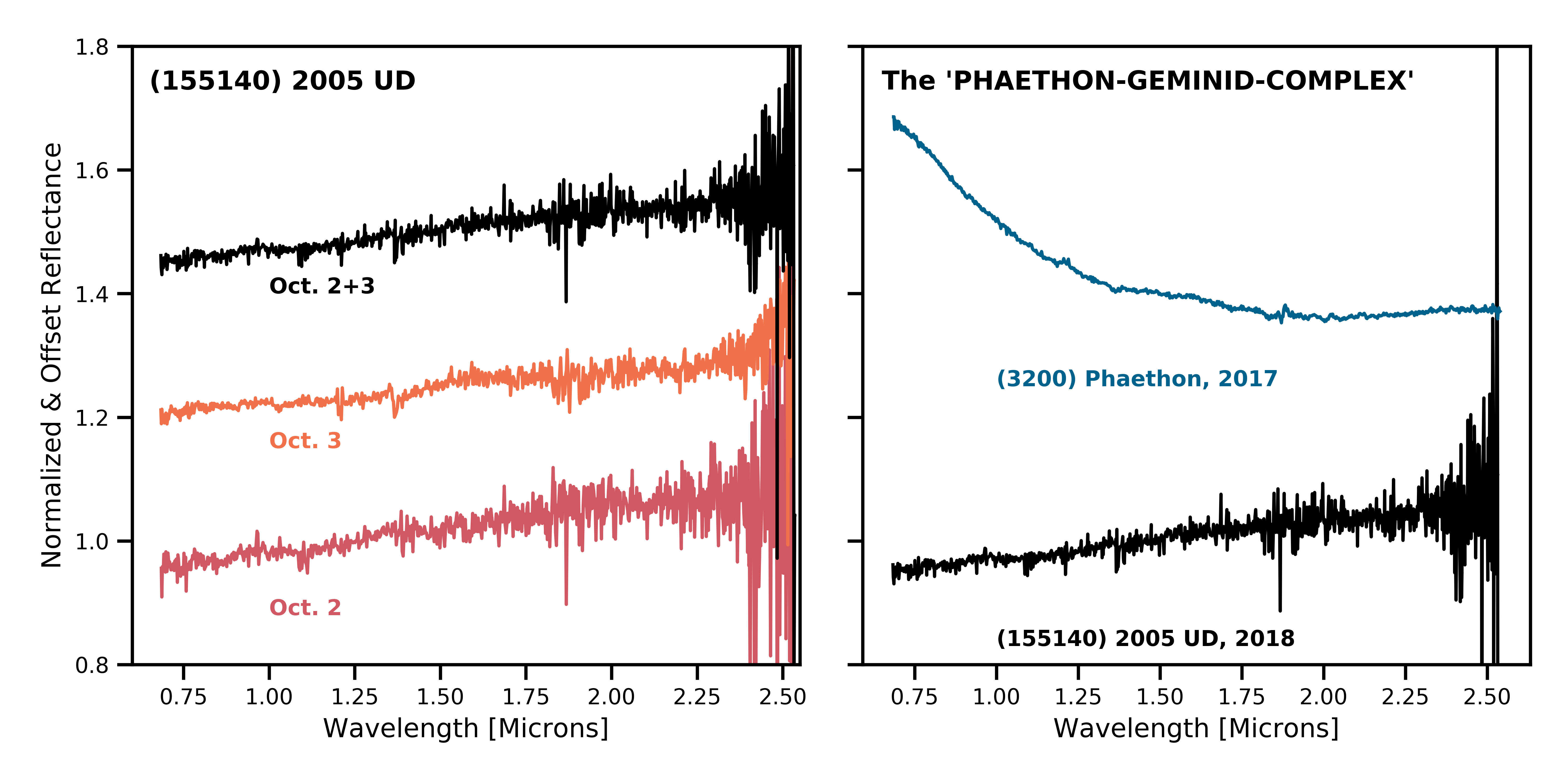}
\caption{Left: The observed near-infrared reflectance spectra of (155140) 2005 UD as observed on October 2 and 3 2018. All spectra are shown to be primarily linear and weakly red-sloped. Right: The combined spectrum of 2005 UD from October 2 and 3 is compared against the average near-infrared reflectance spectrum of its proposed parent body, (3200) Phaethon, from \citep{2018AJ....156..287K}. Dates or object names are listed underneath each spectrum, and all spectra were normalized such that $R(1.5 \mu{m}) = 1.0$ and offset vertically for clarity.  The spectrum of Phaethon has had a long-wavelength thermal emission component removed analytically. Phaethon and UD, despite their widely reported similar visible surfaces, are shown to be highly different at near-infrared wavelengths. A third spectrum taken September 19th of the same year is of much worse quality and is described in the text.}
\end{figure}

The reflectance spectra obtained on October 2 and 3 2018 dates of 2005 UD are presented and compared with the reflectance spectrum of Phaethon from \citealt{2018AJ....156..287K} in Figure 1. The September 19 data is much noisier compared to the other two spectra, so it is only described here and omitted from Figure 1. The reflectance spectrum of 2005 UD weakly red throughout the near-infrared ($\sim0.7-2.5 \mu{m}$ on all three dates, though we note that the use of MORIS as a guider with SpeX can decrease the signal in shortest wavelength region ($<0.85\mu{m}$) due to low throughput near the visible/near-infrared dichroic mirror (Personal Communication, S.J. Bus.) We also note that while the slit was aligned to the parallactic angle, the object's non-sidereal tracking rates were so large that moderate short-wavelength slit losses may be possible. Both of these issues would result in \textit{lost} flux, and thus artificially lower reflectance values in the $<0.85\mu{m}$ region if relevant. The lack of any large change in slope at the shortest wavelengths indicates that neither process had a significant effect on the data.

To check for consistency in spectral slopes between our three nights of observations, we consider the region $0.85-2.4 \mu{m}$ and calculate values for $S'$, where $S'_{\lambda}(\lambda_1,\lambda_2) = (dS/d\lambda) /S_{\lambda}$, of $1.5\pm0.2$, $0.7\pm0.2$, and $0.5\pm0.1 \%/(0.1\mu{m})$ for September 19th, October 2nd and 3rd respectively. The uncertainties are those from the formal linear curve-fitting. The combined October 2 and 3 spectrum has $0.6\pm0.1 \%/(0.1\mu{m})$. For context, the Bus-Demeo average C-type has $S'=1.16\pm0.07$ over the same range, so we find that 2005 UD is less red than a C-type but still red in the near-infrared. We note that \citet{2020ApJS..247...73M} analyzed decades of similar observations and found a systematic $4.2\%/\mu{m}$ uncertainty over this whole wavelength range ($0.8-2.4\mu{m}$), larger than our uncertainties derived from our linear fits. However, those same authors note that regular observations of standard stars (thus requiring regular re-acquisition of the target on the slit) do mitigate much of this systematic uncertainty. We also utilized standard star observations close in airmass ($\Delta X < 0.1$) which is the other dominant factor in this systematic error. The rotation period of 2005 UD is 5.235 hours \citep{2020PSJ.....1...15D}, so our October 2 and 3 observations are separated by almost exactly $\sim4.0$ rotation periods, and the September 19th and October 2nd observations are separated by slightly more than $\sim59$ rotation periods. These three separate nights of observations do not show evidence for the previously reported possible heterogeneous surface of 2005 UD \citep{2007A&A...466.1153K}, but that is perhaps unsurprising as the three sets of observations span the same $\sim\frac{1}{4}$ of the surface. The observing geometry did change significantly, so some difference might have been expected. The redder slope of the September 19th observations is statistically different, but given that the spectrum is likely phase-reddened \citep{2012Icar..220...36S}, was taken at a low enough heliocentric distance that the object likely has thermal emission contaminating the longer wavelengths (see, e.g.  \citealt{2009M&PS...44.1917R, 2003Icar..166..116D, 2018AJ....156..287K}), and is much noisier due to the lower total integration time on a dimmer target, we suspect the difference is not resulting from heterogeneous surface properties. We again note that measuring slope-differences at the level of subtlety discussed here is at or near the limit of what can be provided by traditional observing schemes with this instrument, so our actual knowledge of slope variations on 2005 UD is driven by systematic uncertainties which are challenging to quantify in more detail than we have described in this section.

\section{Phaethon vs. 2005 UD}
The revelation that (3200) Phaethon and (155140) 2005 UD might have very different near-infrared reflectance spectra is very surprising given the ample similarities of the two objects (very similar and uncommon visible reflectivities, similar and uncommon orbits, both are meteor shower parent bodies, etc), so two questions must be addressed: is the difference \textit{real} and if it is, what are plausible mechanisms to explain it?

As for whether or not the difference is real, there are multiple reasons that we believe the results. First, both sets of observations (of UD and of Phaethon) compared in Figure 1 were taken by the same instrument, by the same observers, and reduced using the same packages and methods. Second, due to the large non-sidereal rates which 2005 UD moved at during the portion of its 2018 apparition in which we observed it, we had to use different local telluric stars each night (but the same master solar analog), which when combined with the almost identical retrieved reflectance spectra on each night makes flukes of calibration due to poor choice of calibration stars seem unlikely. Third, though perhaps most important, the retrieved spectrum of Phaethon shown from \citealt{2018AJ....156..287K} is very consistent with many previous observations of the object, and thus the pipeline by which both datasets has been reduced is validated. The difference between the objects seems very real, though we again remind the reader of the possible calibration issues in the 2005 UD data at shorter wavelengths due to the dichroic and parallactic angle issues mentioned previously.

There are then two kinds of scenarios to be considered when attempting to explain these objects discrepant physical properties: either they are genetically/dynamically related somehow, and some process has made their surfaces appear different, or they are not related and their similarities are either coincidental or simply by-products of some larger process in the inner and middle Solar System. 

If the two objects really are only coincidentally similar, then there should be multiple pathways into Phaethon / Geminid - like orbits. While there are serious issues to be resolved in understanding whether or not Phaethon is a heated and devolatilized member of the Pallas Collisional Family \citep{2010A&A...513A..26D, 2018AJ....156..287K,2020arXiv201010633M}, 2005 UD's spectrum is a very poor match for Pallas or any member of the Pallas Collisional Family (they are all strongly blue in the near-infared, many with Phaethon-like concave up spectra), suggesting that those problems would be even more challenging to resolve for it. Furthermore, the surface of the less studied candidate member of the Phaethon-Geminid-Compolex 1999 YC \citep{2008M&PSA..43.5055O} is an even worse fit \citep{2008AJ....136..881K}, so the problems continue in that arena. Dynamical studies of how to implant objects into Phaethon-like orbits from source populations beyond the Pallas family (e.g. other volatile-rich asteroid families, the Outer Main Belt, Hildas, JFCs, etc.) could be extremely useful in answering the plausibility of this scenario. The significant inclination ($\sim22^\circ$) of Phaethon and 2005 UD ($\sim29^\circ$) limits the number of areas to search

If the two objects actually do share some sort of meaningful relationship, then some divergent process should be able to create both kinds of surfaces. We consider the five following processes:

\begin{itemize}
    \item Parent Body Heterogeneity: If the precursor object that disrupted into Phaethon and 2005 UD had variable composition over its surface and throughout its interior (such as if it had been fully or partially differentiated, as has been argued for (24) Themis in \citealt{2010GeoRL..3710202C}), then the daughter fragments could inherit some of these differences and appear different from each other. Considering that Phaethon itself appears to have a largely homogeneous surface \citep{2018AJ....156..287K} and dominates the mass of the two objects combined\footnote[1]{If Phaethon has an average diameter of 5.7 km \citep{2019P&SS..167....1T} and 2005 UD has an average diameter of $\sim$1.3 km \citep{2006AJ....132.1624J}, then Phaethon has $\sim99\%$ of the mass of the combined system if their densities are similar.}, it seems unlikely that UD would somehow inherit different properties by chance. Furthermore, if the Geminids are a by-product or related to the breakup event, then they too should be heterogeneous, which they do not appear to be outside of variable Sodium depletion \citep{2010pim7.conf...42B} often-but-not-exclusively attributed to their low perihelion distance.(In principle, we cannot rule out some stranger scenario where the progenitor object was obviously differentiated or heterogenous, but neither daughter object nor the Geminids inherited any unambiguous heterogeneities by chance.)

    \item Differential Space Weathering: Space weathering is a collection of processes (energetic particle impact, UV irradiation) that acts to change the spectral slope of an object. While in general the process is said to make the reflectivity of a planetary surface redder and darker \citep{1993JGR....9820817P} (true for rocky surfaces like the Moon or S-type asteroids), there is evidence that some volatile rich surfaces like those of CI or CM chondrites might become bluer \citep{2017Icar..285...43L}. Furthermore, recent laboratory \citep{2020Icar..34613775T} and in-situ measurements at (101955) Bennu \citep{2020Sci...370.3660D} suggest that for carbonaceous surfaces, surfaces might \textit{initially} become bluer only to brighten and redden afterwards. It is important to note that traditional space weathering products such as nano-phase iron seen in lunar regolith have not been detected in abundance on CI or CM chondrites and it is expected the same stimuli are resulting in fundamentally different processes on carbonaceous objects, even though all of these effects are labelled 'space weathering'. In general, space weathering apparently has a much larger effect at visible wavelengths that tapers off into the near-infrared, so the very similar visible reflectance spectra of Phaethon and 2005 UD \citep{2006AJ....132.1624J} makes this scenario seem unlikely. Space weathering, thought to decrease strongly in importance with increasing distance from the Sun, must play some role in modifying the surfaces of these two objects. If Phaethon and 2005 UD have different surface effective surface ages, this would allow them to be at different parts of the blue-ing-then-reddening trends found in works like \citet{2020Icar..34613775T} and \citet{2020Sci...370.3660D}. However, the magnitude of changes estimated from those and other works are not enough to account for the observed difference between the two objects.\footnote{One issue complicating an estimation of the role of Space Weathering on these objects is whether or not they are active enough to outpace any potential weathering, e.g. if the weathered material is quickly lost due to radiation pressure or the Solar Wind or any of the other processes proposed for Phaethon.}
    
    \item Grain Size Effects: The distribution of grain sizes on a planetary surface can change the way that light reflects off of it significantly (see, e.g., \citealt{1984JGR....89.6329C}). Moreover, more massive bodies should be able to retain even smaller grains on their surfaces due to their stronger gravity -- and Phaethon is much more massive than 2005 UD. However, Phaethon also seems to be losing the finest ($\sim1\mu{m}$) grains off of its surface due to solar radiation pressure \citep{2010AJ....140.1519J}, so it is challenging to make a-priori estimate of how the two objects might have different grain size distributions. A finer point is that the texture of the larger grains can mimic the appearance of smaller grains should it be of the right scale. That being said, while grain size effects can be complex, there is no reason to expect that these effects have little effect at visible wavelengths and a large effect in the near-infrared. Furthermore, \citealt{2020PSJ.....1...15D} estimated that both objects have grain sizes in the $\sim1-10mm$ range, well outside of any transition in scattering regime for the wavelengths considered. Again, in principle, this may play \textit{some} role in the overall difference of the two surface reflectance spectra, especially if these similarly-sized larger grains have significantly different textures, but it cannot be the dominant cause of the difference.
    
    \item Phase Angle Effects / Phase Reddening: Observations of small bodies at higher phase angles ($>30^\circ$) can appear artificially redder, and this effect is most prominent at near-infrared wavelengths \citep{2012Icar..220...36S}, making it initially seem to be a better explanation than the other processes discussed so far. The two highest quality observations of 2005 UD presented here were at 30.5 and 34.2 $^\circ$ phase, while the comparison observations of Phaethon were at $\sim22^\circ$. This difference of $8-14^\circ$ is enough for only a very slight increase in slope, if any, and thus not enough to explain the difference between the two objects. \citealt{2012Icar..220...36S} found that observations at $\sim30^\circ$ phase had a negligible slope change from those at lower phase angles, from which we infer that any uncertainty in the slope of our combined 2005 UD spectrum is likely dominated by factors other than phase angle effects.
    \item Differential Thermal Alteration: The individual materials that make up a planetary surface can be altered by high temperatures, including physical changes like the break down or exfoliation of boulders (recently studied up close by OSIRIS-REx at the near-Earth asteroid (101955) Bennu, see \citealt{2020NatCo..11.2913M}) and chemical changes brought on in individual minerals \citep{1993Sci...261.1016H, 1996M&PS...31..321H}. The latter is thought to be one of the most important processes that drives spectral diversity among carbonaceous asteroids, particularly among near-Earth objects \citep{2009MNRAS.400..147M}. Phaethon and 2005 UD are both in highly eccentric orbits with very high but distinct perihelion temperatures. If some material could be identified that can produce a Phaethon-like spectrum when heated to a Phaethon-like temperature as well as a UD-like spectrum at a UD-like temperature then their different spectral behaviors could have a natural explanation that could then be studied on other objects.
\end{itemize}
To assess the plausibility of this last scenario, we performed a blind search for high-quality spectral match throughout all of the RELAB database \citep{1983JGR....88.9534P, 2004LPI....35.1720P} after making an initial cut to only include spectra whose visible albedos were less than $15\%$ at $0.55\mu{m}$ to restrict the candidates to those with low albedo while still being agnostic to the discrepant estimates of Phaethon's albedo \citep{2016A&A...592A..34H, 2018AJ....156..287K, 2019AJ....158...97M}. As with previous studies of Phaethon's surface (e.g. \citealt{2007A&A...461..751L}), the best matches of both objects were heated CI and CM chondrites -- both those heated in the laboratory and those found already showing evidence of thermal metamorphism. The in-air spectra of the CI chondrite Ivuna after heating to 873 K and 973 K were among the best fits for Phaethon and 2005 UD respectively, with and without forcing the visible spectrum to be blue-sloped. Both in-air laboratory spectra have linear blue slopes at visible wavelengths, while the 873 K spectrum becomes linear and red near $\sim0.9-1\mu{m}$ and the 973 K spectrum continues as blue and linear in the near-infrared without any very obvious slope break.\footnote{We note for any readers interested in utilizing or inspecting these Ivuna spectra that they are labelled in RELAB in degrees Celsius, not Kelvin, e.g. 600 $^\circ$C and 700 $^\circ$C, and seem to be from \citealt{1996LPI....27..551H}.} This trend with heating, whereby the visible reflectance remains blue but the near-infrared reflectance goes from linear and red to blue-sloped is the right kind of trend to explain these objects. The matches were imperfect in key ways, however, the 873 K spectrum was slightly too red for 2005 UD and the 973 K spectrum was not sufficiently blue to match that of Phaethon, and it did not have the same slight convex-upwards curvature seen throughout the same wavelength range. Assuming an albedo like that of \citealt{2018AJ....156..287K}'s for Phaethon for both objects ($p_v = 0.08 \pm0.01$) and a standard emissivity value ($\epsilon=0.9$), we would expect their surfaces to be heated to approximately these temperatures. Phaethon's surface-averaged temperature $T_{ave}\sim969K$ and sub-solar temperature $T_{SS}\sim1085K$ at perihelion are similar and slightly higher than the 973 K laboratory data respectively, while 2005 UD's temperatures ($T_{ave}\sim895K$, $T_{SS}\sim1002K$) at its perihelion are similar and higher than the 873 K laboratory data as well. The explanation that 2005 UD is simply a less thermally metamorphosed Phaethon is an exciting prospect, but the hypothesis is largely based on a limited number of laboratory measurements that were conducted in-air and not under a space-like vacuum. We deemed it necessary to collect more laboratory data of heated CI chondrites to confirm the spectral trends seen and try to better understand how well they could constrain the thermal histories -- and thus relationship -- between these two objects.

\section{Laboratory Reflectance Spectroscopy} \label{sec:lab}
\subsection{Methodology}
In order to test the hypothesis that Phaethon and 2005 UD are made of the same material but heated to different degrees, new laboratory measurements of the reflectance of CI Chondrites at higher temperatures than available in the literature from RELAB had to be obtained, and ideally in a vacuum as opposed to in-air. In particular, the heating experiments should be designed in such a way to both mimic the grain sizes thought to be relevant for Phaethon and UD ($\sim 1-30 mm$, see \citealt{2020PSJ.....1...15D}) and the vacuum of space. To achieve this goal, we constructed a vacuum heating chamber (affectionately referred to as the "Bar-B-Cube") in the Reddy Spectroscopy Lab at the University of Arizona. After initial sample preparation, which might vary based on the sample and topic being investigated (e.g., grinding and size sorting with a mortar and pestle), the sample is placed inside an No. 200 Nickel sample cup (chosen based on Nickel's high melting point and ease with which it can be machined) and lowered into the chamber from above. The sample cup is placed on the center of the heating element and the sapphire window on the top of the chamber is tightened. The vacuum pump is then turned on and the chamber is slowly lowered to $\sim10^{-6}$ Torr or lower. The vacuum chamber was able to reach pressures as low as $\sim10^{-8}$ Torr if untouched for several hours. Highly volatile rich samples, such as CI or CM chondrites, require this process to be quite slow as to avoid sudden outgassing disrupting the surface of the sample once the pressure inside approaches the triple point of water at room temperature ($\sim7$ Torr). Reflectance spectra of the sample within the chamber are obtained by first calibrating on a nearby spectral standard through an identical sapphire window and then moving the optical fibers to observe the sample. The optical fibers are held in a standard orientation for both the calibrations and the science observations using a 3D printed part which keeps the two fibers $30^{\circ}$ apart. The fibers are held in a near-vertical orientation otherwise in both configurations to avoid differential flexure which might introduce subtle artifacts. This procedure was successfully tested on a variety of meteoritical and artificial samples to verify that the retrieved spectra were consistent with those measured through standard in-air procedures, though the spectra are often somewhat noisier. We note that we used a lamp optimized for the near-infrared, so the reported spectra are only shown beyond $0.5\mu{m}$. While the retrieved spectra are consistent in spectral behavior with their in-air counterparts after calibrations, the absolute albedo (vertical scale) is likely not after heating for volatile-rich samples. As volatiles leave the samples, the surface subsides somewhat in a way that cannot be measured well enough to correct for the change. As a result, the reflecting surface is further away from the optical fibers and thus appears artificially darker to the spectrometer.

The heating of a particular sample is done in increments of $50-150 K$ to both attain a fine temperature resolution to record spectral changes over and also not to cause the sample to lose too many volatiles at once and risk exploding out of the cup into the vacuum. The general procedure is to raise the temperature of the heating element to a desired temperature, wait until the temperature stabilizes at the set value ($~5-15$ minutes depending on how high the temperature is set), and then keep the temperature at that level for 1 hour or until outgassing of the sample has been decreasing steadily for 15 minutes, whichever is longer. For the volatile-rich samples tested, this was almost always 1 hour with the exception of temperatures near $\sim 950-1000$ Kelvin, where outgassing from the sample was much more significant. (This is also the highest temperature range reached for the previous study of Ivuna \citealt{1996LPI....27..551H}.) Once this time limit is reached, the sample is allowed to cool radiatively back to near-ambient temperatures ($\sim 300$ K), which could take up to 2 hours from the highest temperatures the samples were heated to. Blackbody emission clearly contaminated the retrieved spectra at any higher temperature than room temperature, and all reflectance spectra suspected of having a blackbody component were discarded when possible. A full heating, cooling, and observation cycle took approximately $\sim 3$ hours on average. The maximum temperature reliably maintained by the heating element with a filled sample cup was $\sim 1350$ K, or approximately $\sim 375$ K higher than the samples of Ivuna we compared our data to from RELAB.

\subsection{Heating CI Orgueil}
This study largely focuses around our heating of a sample of the CI Chondrite Orgueil, specifically the sample 1070B from the Vatican Observatory's Meteorite Collection. The previous CI chondrite that was heated (Ivuna, heated up to 973 K in \citealt{1996LPI....27..551H}) has somewhat different reflective properties than Orgeuil, and our sample preparation routine was not identical to theirs, so differences are to be expected. In particular, Ivuna is somewhat bluer at near-infared wavelengths than Orgueil (at least before heating) and \citet{1996LPI....27..551H} heated their samples as unground chips and then ground them down into a semi-fine powder ($<125\mu{m}$.) Their spectra were also obtained in-air, as opposed to a near-vacuum more analogous to the conditions on planetary surfaces. \citealt{2011Icar..212..180C} found that even slight changes in particular mineral abundances, particularly magnetite, can cause comparatively large slope changes in CI chondrites. For comparison with that study and to meet our goal of matching the estimated grain size for these bodies as best as possible considering the limited amount of sample of Orgueil provided ($\sim0.5$ grams), we divided our sample roughly in two into an unground $\sim 1 mm$ subsample and a $<45\mu{m}$ subsample, which largely stuck together in larger clumps after being sorted into the sample cup. While the orbital history of the CI Ivuna parent meteoroid is not known, the orbital history of the Orgueil strongly suggests that the parent was on a Jupiter Family Comet or Halley Type Comet orbit \citep{2006M&PS...41..135G}. The nature of the CI chondrites in the context of their parent bodies is discussed further later. In order for a particular laboratory spectrum of Orgueil to be considered good evidence that a CI-like meteorite could fit the spectrum of the small body in question, the deviation from the fit should therefore be within the range of variability expected from changes in grain size or mineral content (within reason). The reflectance spectra of large mm-sized pieces of the CI Chondrite Orgueil heated from room temperature up to $\sim1350$ K are shown in Figure 2.

\begin{figure}[ht!]
\plotone{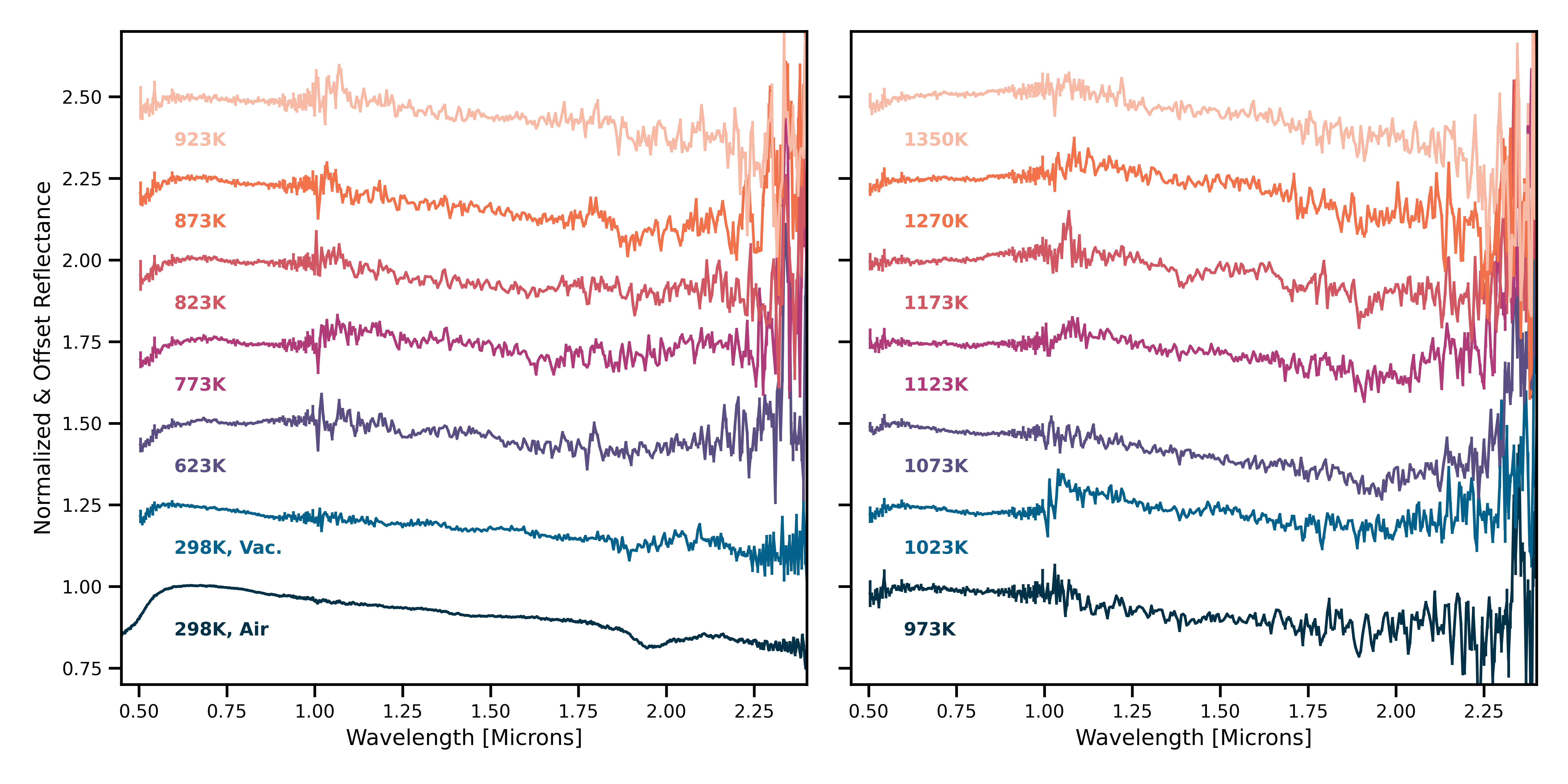}
\caption{The retrieved reflectance spectra of the CI Chondrite Orgeuil after heating to successively higher temperatures are shown, with room temperature to 923 K on the left side and 973 to 1350 K on the right side. Temperatures are listed underneath each spectrum at $0.6 \mu{m}$, and all spectra were normalized such that $R(0.6 \mu{m}) = 1.0$ and offset vertically for clarity.}
\end{figure}

The spectrum of the large grains ($\sim$ mm-sized) of CI chondrite Orgueil is shown to change significantly as it is heated to successively higher temperatures, but the trends observed are not identical to those demonstrated for the CI chondrite Ivuna in \citep{1996LPI....27..551H}. At visible wavelengths, the slope of the reflectance spectra become increasingly linear with increasing heating (a decrease in near-UV absorption strength, as previously noted for Ivuna as well) and has variations in absolute slope while still being blue (negative) until 1123-1173 K, where the slope becomes red and linear. At near-infrared wavelengths, the slope is always overall blue (negative) though curvature (concave-up) is apparent at a handful of temperatures, most prominently between 973 - 1173 K, or approximately the range of temperatures expected for Phaethon's surface at the present day and in the recent past near perihelion. Figure 3 shows a comparison of our Orgueil and \citet{1996LPI....27..551H}'s Ivuna both heated to 973 Kelvin compared with the reflectance spectrum of Phaethon. Figure 4 shows a comparison of the derived spectral slopes for the whole dataset as a function of temperature at a variety of wavelength ranges again compared with Phaethon and 2005 UD. The relationship between the visible and near-infrared slopes, and specifically how they change in the $0.8-1.0 \mu{m}$ region and the curvature at longer wavelengths could indeed be a useful way to identify thermal metamorphism in Orgueil-like bodies, similar to that shown in the in-air Ivuna data of \citealt{1996LPI....27..551H} and suggested in other contexts by \citealt{2017AJ....153...72V, 2016A&A...586A..15M}. The reflectance maximum seen near $\sim0.6\mu{m}$ decreases in prominence with increasing heating, but does not appear to totally disappear as reported for the Ivuna samples in \citealt{1996LPI....27..551H}. We see no evidence for any curvature in the $\sim1.3\mu{m}$ region as was seen in Ivuna, and in fact the slope changes due to heating appear less obvious there. We note that the higher SNR in the room-temperature in-vacuum spectrum appears to be primarily due to the fact that the sample shrunk in size afterwards due to loss of volatiles. In other words, the room-temperature spectrum shown had the sample surface be physically closer to the fiber that was going to the spectrometer, resulting in higher throughput.

Part of the difference between our work on Orgueil and previous work on Ivuna is almost certainly the usage of finer grain sized powders in \citep{1996LPI....27..551H}, which \textit{generally} would result in redder spectra, but the overall trends are different in a way that hints at more processes in play. This is perhaps to be expected, as carbonaceous chondrites are often highly heterogenous in composition at small scales \citep{2006mess.book..679B}. Another critical difference is that our data were obtained in-vacuum, and thus in an environment more similar to that of Phaethon and UD, as opposed to the in-air spectra presented of Ivuna. 

While the spectra of Orgueil heated to Phaethon-like temperatures show many similarities with Phaethon's reflectance spectrum (with some key differences discussed later), the fact that none of the retrieved spectra show red or neutral slopes in the near-infrared seem to conflict with any positive match with 2005 UD. Even if the standard is broadened to 'a near-infrared slope that is redder than the visible slope of the same sample', there seem to be no retrieved spectra that meet the criterion.

\section{Further Discussion}
\subsection{Spectral Comparisons}
In this study, we have obtained the first near-infrared reflectance spectra of the proposed fragment body of (3200) Phaethon (155140) 2005 UD and found the two objects to be strikingly different. Phaethon's reflectance spectrum is concave-up and overall blue-sloped throughout the near-infrared, while UD's is approximately linear and very weakly red-sloped. One process that we hypothesized could explain their discrepant spectral slopes and curvatures would be thermal metamorphism due to their highly eccentric but distinct near-Sun orbits, meaning that they were made from the same material originally but had subsequently been altered to different extents. A comparison of our newly obtained data of an increasingly heated sample of CI Orgueil and similar data from \citet{1996LPI....27..551H} of CI Ivuna data to (3200) Phaethon's properties is shown in Figure 3.

\begin{figure}[ht!]
\plotone{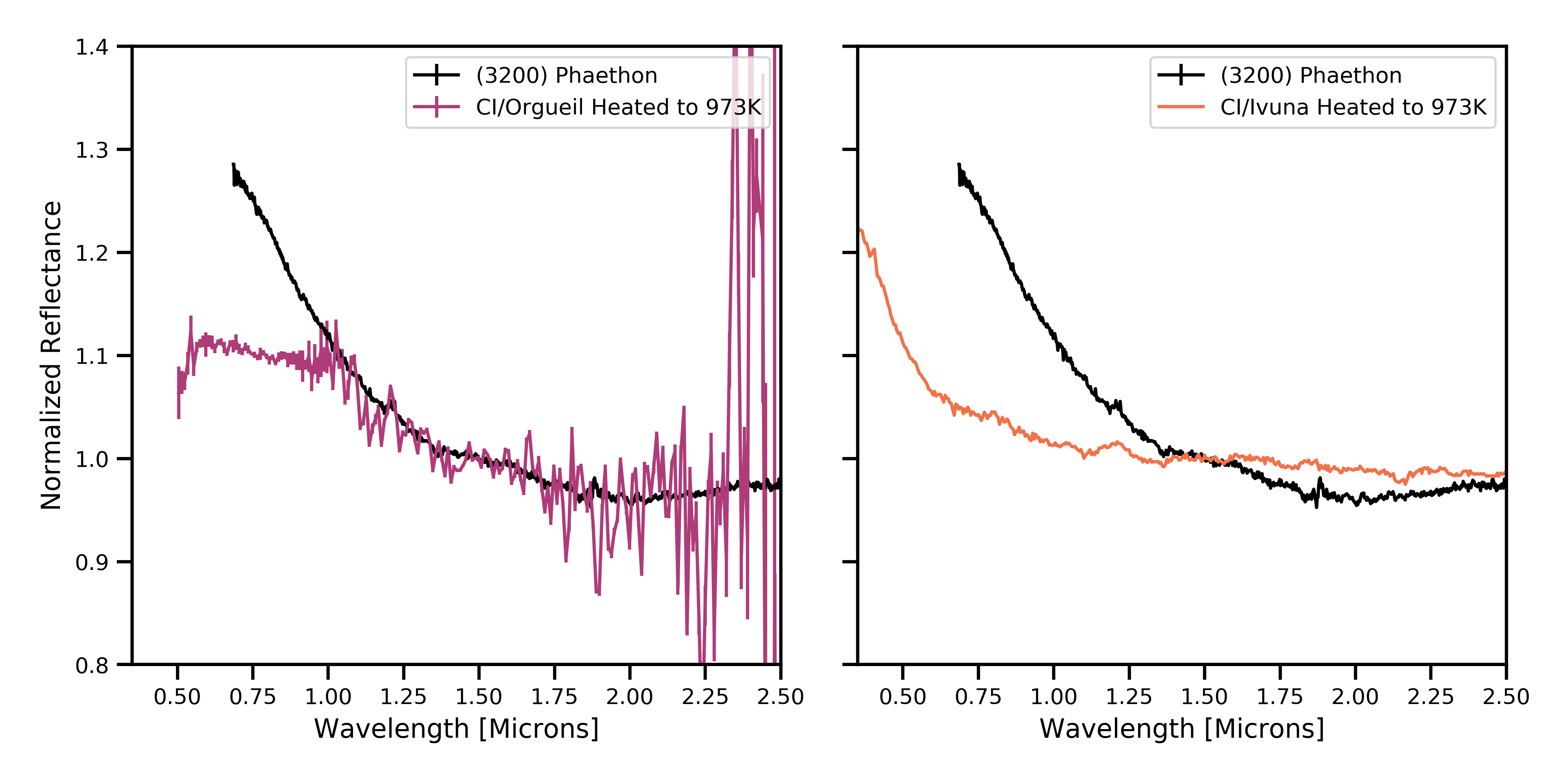}
\caption{The reflectance spectrum of (3200) Phaethon (black) compared with our sample of the CI Orgueil heated to 973 Kelvin (left, maroon) and \citealt{1996LPI....27..551H}'s CI Ivuna heated to the same temperature (right, orange), all normalized such that $R(1.5\mu{m})=1.0$. Neither are perfect fits to the overall spectrum, but both show qualitative similarities that might indicate some material similarity given the lack of known samples of CI chondrites on the ground.}
\end{figure}

Through visual inspection alone of Figures 2 and 3, it can be seen that none of the heated samples appear similar to 2005 UD at near-infrared wavelengths, while those in the temperature range expected for Phaethon (roughly $1000-1200$ K) are indeed approximately linear and blue-sloped at visible wavelengths while also being blue and concave-up at near-infrared wavelengths in good qualitative agreement. Phaethon shows no near-UV decrease in reflectance above $0.35-0.4 \mu{m}$, unlike the data for Orgueil shown here. If the near-UV drop-off in our data is real (and not driven by, perhaps, very low light levels and a light source optimized for the near-IR), then we conclude that Phaethon's spectrum is better described as Ivuna-like a short wavelengths and Orgueil-like at longer wavelengths. One other key aspect to note is where the VIS/NIR slope change occurs. In our data on Orgueil, the slope change occurs somewhere between $0.8-1.0\mu{m}$ but can't be discerned more clearly due to its proximity to a changeover in detector usage in our spectrometer near those wavelengths. In the Ivuna dataset of \citealt{1996LPI....27..551H}, the slope change occurs nearer to $0.6-0.7\mu{m}$. In Phaethon and 2005 UD, the location of this spectral slope change is near to $\sim0.8\mu{m}$ but, again, is often the area where to separate datasets (one from a visible-wavelength instrument and one from a near-infrared instrument) are stitched together. Again, the Orgueil dataset seems 'closer' than Ivuna does to matching Phaethon, but the location of the spectral slope change needs to be tied down more tightly with future work. Neither Orgueil nor Ivuna fits 2005 UD well, but a combination thereof seems to share many similarities with Phaethon. The rarity of CI Chondrites is likely a contributing factor to the existence of a more "perfect" match at this date.

In Figure 4, we show a comparison of the derived slopes S' \citep{1990Icar...86...69L} of the laboratory data at three wavelength ranges ($0.60-0.74\mu{m}$, $1.1-1.60\mu{m}$, and $1.8-2.2\mu{m}$) compared against the slopes of Phaethon and 2005 UD calculated over the same wavelengths. These ranges were chosen to try to encapsulate the changing curvature of the spectrum while avoiding the noisier parts of the laboratory data where there was a change in detectors. Where the laboratory data intersects the colored areas corresponding to each NEO is where their slopes agree (within errors) at those wavelengths. In agreement with visual inspection, Phaethon's spectrum appears most similar to the Orgueil samples heated to $973-1073$ Kelvin. 2005 UD's apparent argreement in slope at longer wavelengths is not indicative of an actual spectral \textit{match}, but instead due to the increasing curvature of the sample of Orgueil throughout the near-infrared. The $1.1-1.60\mu{m}$ slope shows much less variability than the other two regions, with slopes near $S'\sim-1.25$ to $-1.50$ for most temperatures. The $0.6-0.74\mu{m}$ region generally blues from $\sim623K$ to $\sim1073K$ before sharply becoming more red, and a similar trend is shown for the $1.8-2.2\mu{m}$, though the bluest slope is achieved at $\sim973K$. The fact that these changes are centered around the $973-1073K$ region is unsurprising, as many materials common in CI chondrites (namely phyllosilicates) begin to break down at these temperatures into other substances. Despite the fact that the laboratory data appears to intersect the spectral slopes of Phaethon for all three spectral regions within the range of temperatures expected for this body, we argue that this is indicative of a similarity between the sample and the NEO as opposed to a direct spectral match considering the imperfections in the comparison detailed in the previous paragraph as well as Figure 3.

\begin{figure}[ht!]
\plotone{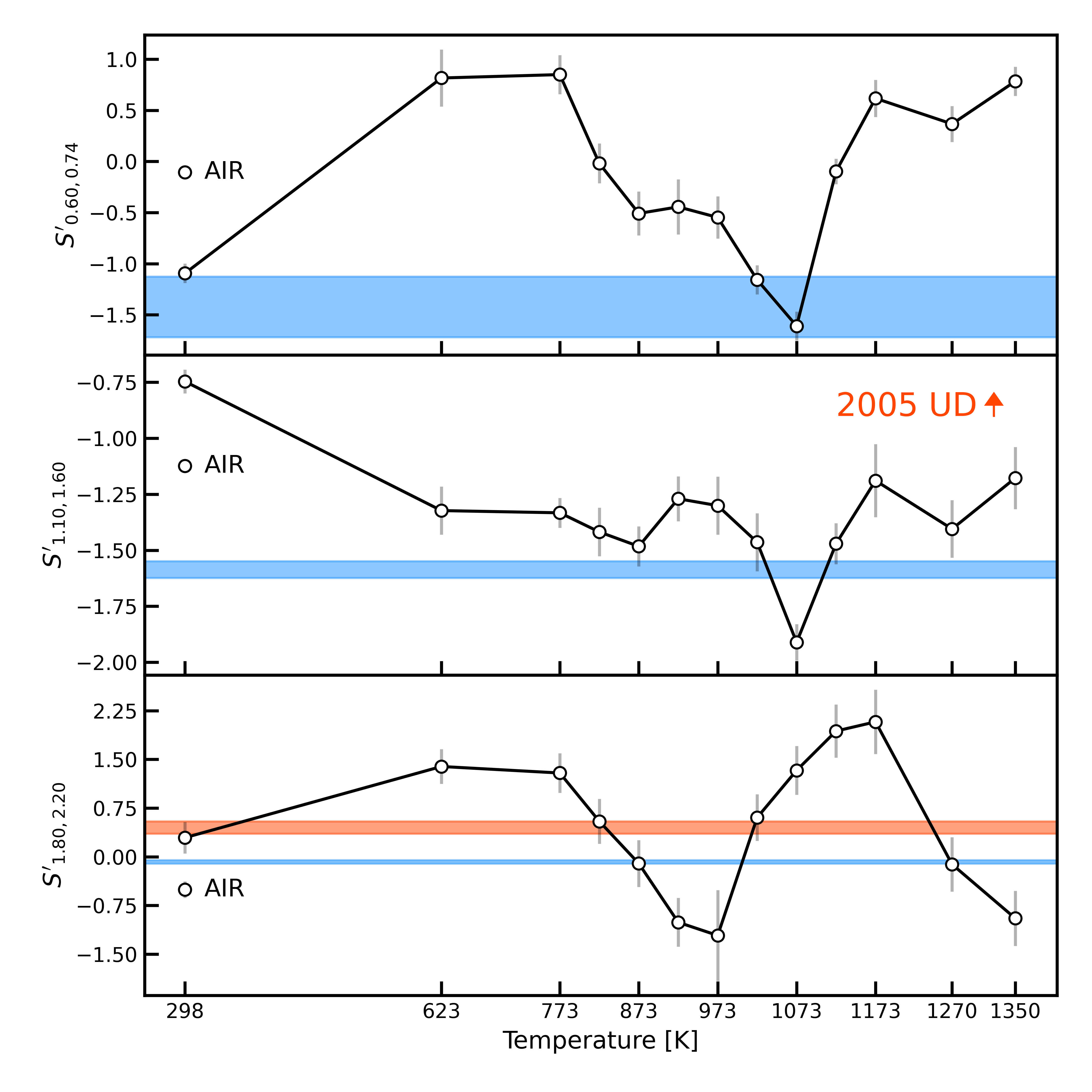}
\caption{The derived spectral slopes (black, in the $S'$ notation of \citet{1990Icar...86...69L}) at visible ($0.60-0.74\mu{m}$) and near-infrared ($1.1-1.60\mu{m}$ and $1.8-2.2\mu{m}$) of our sample of CI/Orgueil as a function of maximum temperature reached. The spectral slopes are compared with those of Phaethon (in blue, using the spectrum from \citet{2018AJ....156..287K}) and 2005 UD (in orange-red).}
\end{figure}

\begin{figure}[ht!]
\plotone{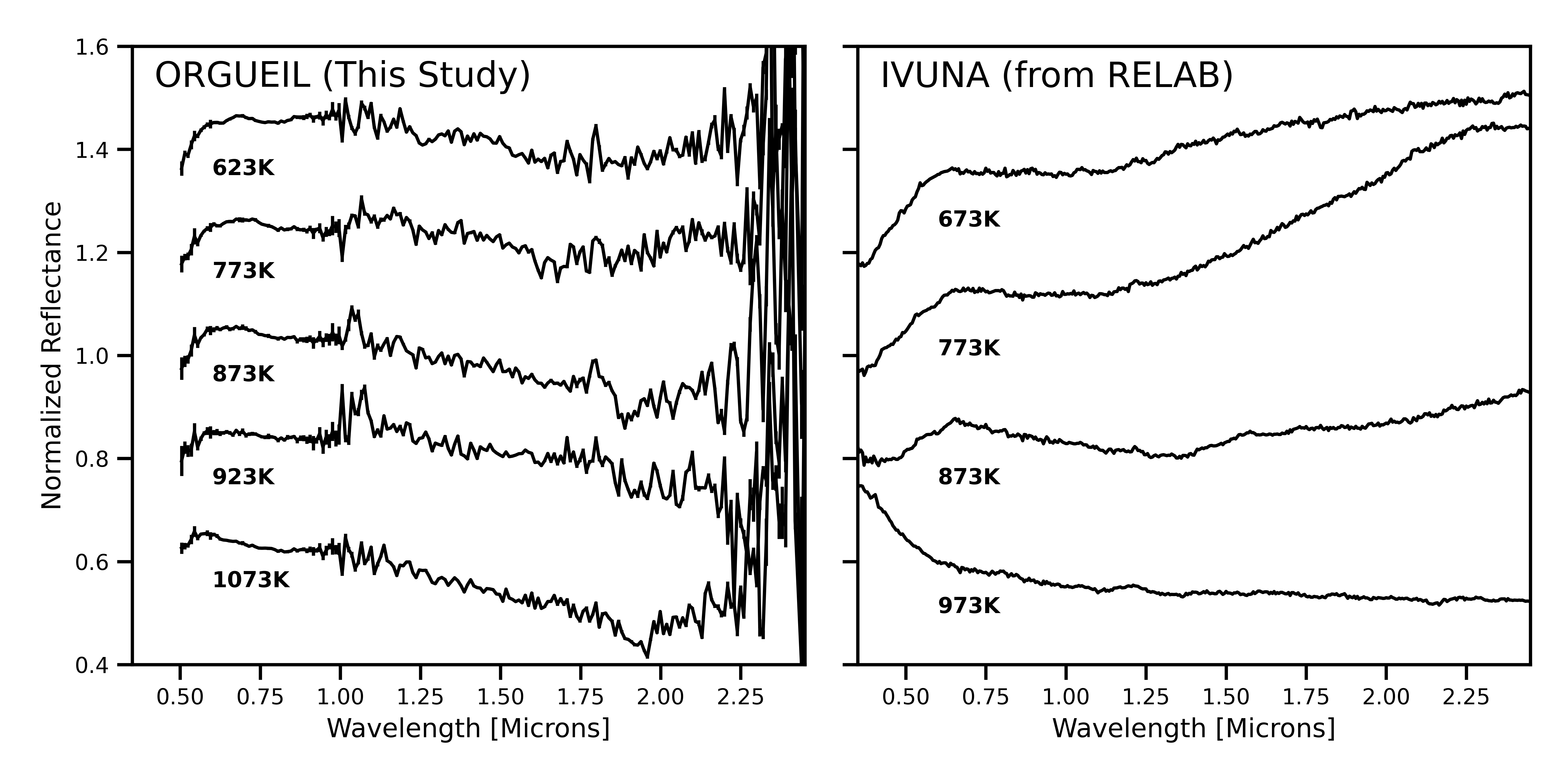}
\caption{
A comparison of the new laboratory data of CI/Orgueil from this study (left) with the RELAB samples of CI/Ivuna (right, \citealt{1996LPI....27..551H}). The spectra shown are subsets of both datasets centered around the temperatures most relevant for the study of 2005 UD and Phaethon. The spectra were all normalized at $0.6\mu{m}$ and subsequently offset vertically to facilitate comparison with Figure 2. The process by which each of these sets of spectra were produced and reduced into their current forms is not identical, see the text for details.
}
\end{figure}

In other words, our new laboratory data and pre-existing laboratory data support the hypothesis that Phaethon's surface resembles that of a CI Chondrite-like material that has been heated significantly due to its orbit. However, the same cannot be said for 2005 UD. While a red slope in the NIR and a blue-slope in the visible can be achieved with the Ivuna data of \citealt{1996LPI....27..551H}, the change in slope is at altogether the wrong location, and no red-sloped spectrum is ever seen in our Orgueil dataset. In summary, we conclude that a heated CI Chondrite origin for Phaethon is compatible with the laboratory data, while the same does not appear to be true for 2005 UD given the currently available spectra.

\subsection{The Relationship Between Phaethon and 2005 UD}
In Section 2, we argued that parent body heterogeneity, differential space weathering, grain size effects, and phase reddening are each unlikely to be able to explain the different spectra of Phaethon and 2005 UD. If the objects were related, then we argued that their spectra could be resultant from differential heating of their surfaces due to their different orbits, but our laboratory data does not appear to support that hypothesis. 

While it is possible there may be some carbonaceous material (either not yet recognized or not available among terrestrial sample collections) for which this hypothetical scenario could have more traction, the fact that CI Chondrites like Orgueil and Ivuna can reproduce all the key features of Phaethon's visible and near-infrared spectrum  while not reproducing 2005 UD's spectrum suggests that the difference might be more than evolutionary and instead related to different origins for the two bodies. 

While the orbital relationship between the two bodies is indeed debated (see, e.g., \citealt{2019MNRAS.485.3378R}), the objects are similar in so many ways that a shared origin is often assumed in spite of contrary orbital evidence. Blue-colored objects are fundamentally rare in the inner Solar System (perhaps $\sim$ 1 in 23 objects, see \citealt{2002Icar..158..146B}), and inactive meteor shower parent bodies even more so. The other best-studied inactive parent body, (196256) 2003 EH1, has a more common organic-rich red surface and a typical cometary orbit \citep{2015AJ....150..152K, 2021PSJ.....2...31K}, both of which lend themselves to more easy explanations than the situation of Phaethon, 2005 UD, and the Geminids. In the absence of other unknown processes operating on these bodies now or in the recent past, we conclude that the least complicated interpretation of the data is simply that the objects are only similar by chance. Should they choose to do so, the $DESTINY^{+}$ team \citep{2018LPI....49.2570A} could thus visit two meteor shower parent bodies that have undergone similar processes yet discrepant origins, thus facilitating a greater understanding of near-Sun objects and meteor shower creation in general, as opposed to studying one interconnected system in great detail. Of course, $DESTINY^{+}$ could also find more compelling evidence for the relationship between the two bodies than we have found here! Future characterization of both bodies is an incredibly compelling task, both before and after they are characterized in-situ.

\subsection{The Origin of the Geminids?}
While an unexpected by-product of trying to explain the difference in spectral properties between Phaethon and 2005 UD, we have also provided more evidence in this study that Phaethon's surface is consistent with that of a heated CI chondrite from the near-UV through to at least $2.5 \mu{m}$. This is consistent with the composition of the Geminid meteors themselves \citep{2005A&A...438L..17K, 2010pim7.conf...42B, ABE2020105040}. Considering that the CI chondrites have been suggested as derived from comets, could the story just be as simple as "Phaethon really is a dormant comet, and the Geminids are just a normal meteor shower"? As appealing as that prospect might be, the problem of Phaethon's modern non-cometary orbit remains, even if its other properties could likely be squared with an outer Solar System origin. (There is also still the question of how best to interpret the origins of the CI chondrites, as while many of their properties are consistent with comets, the existence of hydrothermally altered minerals is outside of the expected material properties of comets significantly.) We consider two scenarios to explain the given telescopic, meteoric, dynamical, and laboratory data.

The first scenario is that Phaethon's progenitor region or parent in the Main Belt was also made out of CI Chondrite like material, but instead of being recently scattered in from the outer Solar System had been emplaced there through a 'Grand Tack'-like scenario \citep{2011Natur.475..206W} billions of years ago. Considering the fact that outer Solar System material has to be in the Main Belt, and a large fraction of the NEOs have to come from there (see, e.g. \citealt{2002Icar..156..399B, 2002aste.book..409M}), there almost certainly have to be members of the NEO population which do have CI-like properties. However, given Phaethon's non-zero inclination and high eccentricity, the number of plausible parent bodies or families is not extremely numerous. The recent work by \citealt{2020arXiv201010633M} shows that the high-inclination inner Main Belt Svea family might be a better match than previous candidates like the family of (2) Pallas \citep{2010A&A...513A..26D}, so further characterization of objects in that family is of interest.

Another scenario is motivated by dynamical studies of the Geminid meteor themselves. First raised in \citealt{1985AVest..19..152L} and recently discussed further in \citealt{2016MNRAS.456...78R} is the idea that the Geminid forming event changed the orbit of the parent body significantly. This particular aspect of the stream forming event is thought to be necessary to explain particular aspects of its duration and spatial extent in the modern day given its young age. If Phaethon's orbit did change significantly around the time of Geminid formation, then all previous backwards orbital integrations of the body likely aren't capturing what \textit{really} happened, and as a result different past dynamical scenarios are possible. In addition to possibly opening up new possible source regions (other parts of the Main Belt, the Jupiter Family Comets, etc.), it could also change how we interpret the 'similar' orbits of 2005 UD and Phaethon in the modern day. If even a rough estimate of how much the Geminid parent orbit would need to change were available, then a suite of dynamical integrations could be investigated to see how much of our current understanding of Phaethon's orbital history is valid. Given the chaotic nature of the inner Solar System and fairly small Minimum Orbital Intersection Distance (MOID) with the Earth ($<0.02 AU$, JPL Horizons), we suspect even a small change in Phaethon's orbit could result in a large change in its modeled dynamical history, and thus a large change in our understanding of the history of the object.

\section{Summary}
The relationship between (3200) Phaethon, the (semi-)inactive parent of the Geminids meteor shower, and (155140) 2005 UD, the inactive parent of the Daytime Quadrantids, is of great interest for several reasons. The bodies appear to have similar visible reflectivities and orbits, neither object is either active in a cometary way or looks obviously like a dormant comet, and they are the primary and possible secondary targets of the upcoming JAXA \textit{DESTINY+} mission scheduled for the late 2020s. In this work, we have synthesized new telescopic observations and new laboratory data to attempt to better understand the relationship between these two objects and their individual properties.

\begin{itemize}
    \item We obtained the first near-infrared reflectance spectrum of 2005 UD utilizing three nights of observations in September and October of 2018, each time finding a primarily linear red-sloped spectrum. Phaethon's spectrum is blue and concave-upwards throughout the same wavelength range (even utilizing the same instrument and reduction scripts), indicating that the difference between the two objects is real despite their very similar properties at shorter wavelengths.
    \item Either the two objects have similar properties by chance or some process or effect has altered them from some common starting point. We make arguments against parent body heterogeneity, differential space weathering, grain size effects, and phase phase reddening as likely origins of the different spectra in Section 2.2, but argue that differential thermal alteration could be a plausible method. The two objects are in highly eccentric near-Sun orbits with surface temperatures where many likely surface materials begin to thermally alter and decompose drastically, and Phaethon's surface (and the Geminids!) has been previously been interpreted as showing signs of previous or ongoing intense heating. CI Chondrites that have been heated to Phaethon/UD-like temperatures seemed promising, but the existing laboratory data did not extend to high enough temperatures to assess the likelihood of this scenario.
    \item We heated a sample of the CI Chondrite Orgueil from the Vatican Meteorite Collection to incrementally higher temperatures up to 1350 Kelvin with a novel laboratory apparatus described in detail in Section 4. We measured the reflectance spectra from $\sim0.5\mu{m}$ to  $\sim2.3\mu{m}$ at each stage after the sample had cooled back to ambient temperatures to see if our hypothesis that a heated CI chondrite could match the features of both Phaethon's and UD's spectrum. Our heating sequence of the CI Orgueil is similar in some ways and different in others from the heating sequence of the CI Ivuna of \citealt{1996LPI....27..551H}, again described in more detail in the text. The spectrum of Phaethon shares many characteristics with both CI chondrites (though neither are a perfect match), but UD is not a high quality match to either object, which argues against the differential thermal alteration hypothesis.
    \item We thus propose that the two objects only have similar spectra by chance, though similar processes likely have acted upon them both. Excitingly, DESTINY+ might be able to visit two unrelated-but-similar meteor shower parents, or it might find evidence for their relationship that isn't easily attainable from remote sensing. We also highlight several areas of future inquiry to better understand Phaethon's modern properties motivated by its apparent similarity to multiple CI Chondrites and recent studies of the Geminid meteors themselves.
\end{itemize}

\acknowledgments
The authors wish to recognize and acknowledge the very significant cultural role and reverence that the summit of Mauna Kea has always had within the indigenous Hawaiian community.  We are most fortunate to have the opportunity to conduct observations from this mountain.

This work was supported by a NASA Near-Earth Object Observations (NEOO) program grant NNXAL06G (PI: Reddy).

We are also thankful to the Vatican Observatory Meteorite Collection for being willing to loan us sample 1070B of the CI chondrite Orgueil, especially for these destructive analyses.

\vspace{5mm}
\facilities{IRTF (SpeX, MORIS)}


\software{NumPy \citep{harris_array_2020}, SciPy \citep{virtanen_scipy_2020}, AstroPy \citep{astropy_collaboration_astropy_2013,astropy_collaboration_astropy_2018}}


\appendix

This appendix provides labelled images and descriptions of our laboratory apparatus such that the reader can better understand its usage, both for a more thorough understanding of this work and in case they one day pursue a similar experimental program.

\section{Photos of the Vacuum Heating Chamber and Samples}
\begin{figure}[ht!]
\plotone{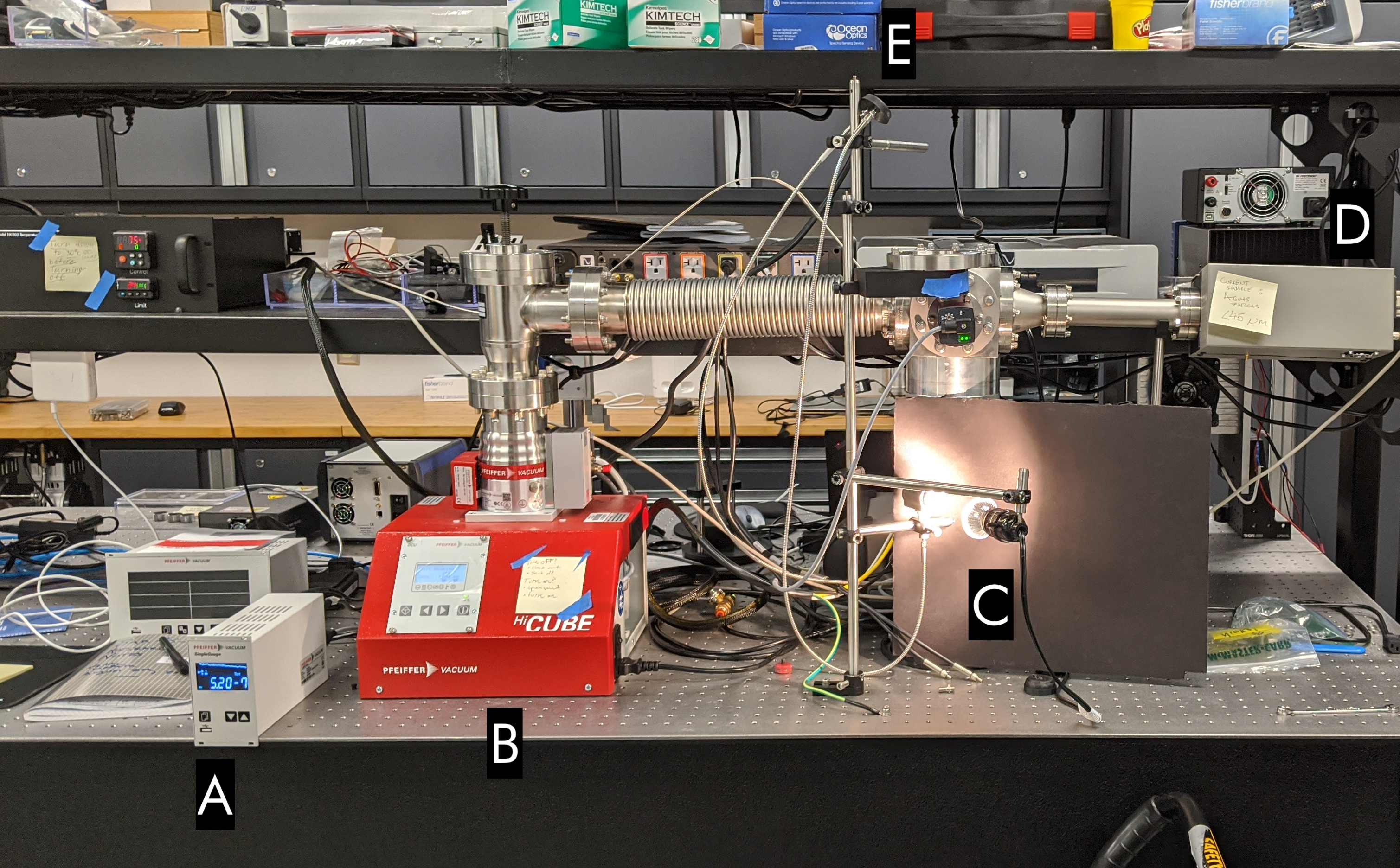}
\caption{The vacuum heating chamber set up ("the Bar-B-Cube") on a laboratory bench at the University of Arizona with individual components labelled. A: the vacuum pressure gauge, which shows the current pressure inside the chamber as well as a graphical summary of recent pressure changes. B: the vacuum pump. C: the light source (a common heating bulb) shining into the input of an optical fiber which is connected to E. D: a residual gas analyzer mass spectrometer, attached to diagnose leaks during the initial testing phase. E: the output of the light emitting fiber as well as the input of the fiber that leads to the VISNIR spectrometer held in a constant orientation with a 3-D printed black ABS plastic holder.}
\end{figure}

\begin{figure}[ht!]
\plotone{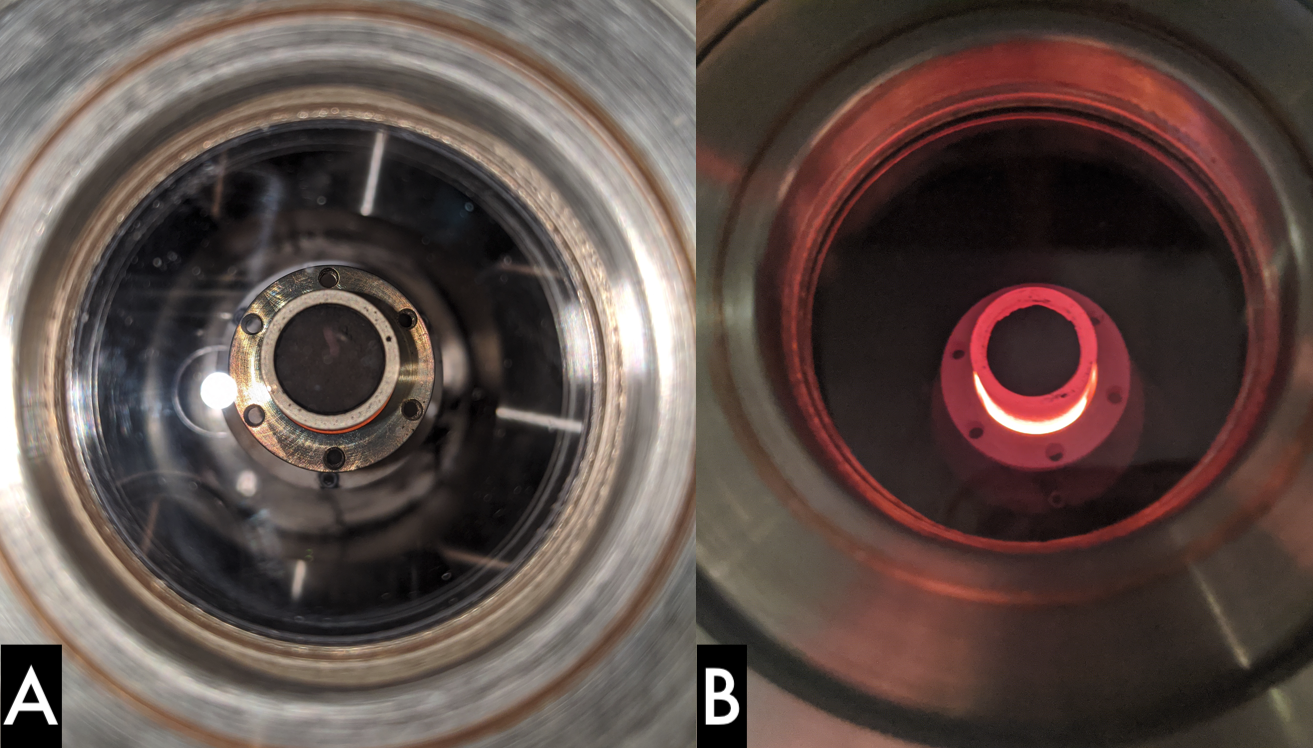}
\caption{A: A close-up view from above the chamber looking through the sapphire viewing window into the interior, showing a filled sample cup ($D\sim0.5"$ for scale) on top of the circular heating element. "Below" the picture is an identical sapphire window mounted at the same height inside of which a spectral standard is mounted. The 3D printed fiber holder fits on top of the window blocking stray external light, and reducing internal reflections. B: Similar to the A, but showing the thermal glow of the Nickel sample cup that was obviously visible at the highest temperatures reached. The bright white-yellow crescent is the edge of the underlying heating unit which is slightly larger in area than the footprint of the sample cup.}
\end{figure}


\bibliographystyle{aasjournal}



\end{document}